\documentclass[letterpaper]{article} 
\usepackage{aaai2026}  
\usepackage{times}  
\usepackage{helvet}  
\usepackage{courier}  
\usepackage[hyphens]{url}  
\usepackage{graphicx} 
\usepackage{natbib}  
\usepackage{caption} 
\usepackage{threeparttable} 
\usepackage{multirow}
\usepackage{booktabs}

\usepackage{algorithm}
\usepackage{amsmath}
\usepackage{algorithmic}
\usepackage{amssymb}
\usepackage{cases}
\usepackage{arydshln}
\usepackage{newfloat}
\usepackage{listings}
\DeclareCaptionStyle{ruled}{labelfont=normalfont,labelsep=colon,strut=off} 
\floatstyle{ruled}
\newfloat{listing}{tb}{lst}{}
\floatname{listing}{Listing}
\newcommand{\ie}{\textit{i}.\textit{e}., }
\newcommand{\eg}{\textit{e}.\textit{g}., }

\title{Unsupervised Motion-Compensated Decomposition for Cardiac MRI Reconstruction via Neural Representation}
\author {
    Xuanyu Tian\textsuperscript{\rm 1,\rm 2},
    Lixuan Chen\textsuperscript{\rm 3},
    Qing Wu\textsuperscript{\rm 1},
    Xiao Wang\textsuperscript{\rm 1},
    Jie Feng\textsuperscript{\rm 4},
    Yuyao Zhang\textsuperscript{\rm 1},
    Hongjiang Wei\textsuperscript{\rm 4}\thanks{Corresponding author.} 
}
\affiliations {
    \textsuperscript{\rm 1} School of Information Science and Technology, ShanghaiTech University, Shanghai 201210, China\\
    \textsuperscript{\rm 2} Lingang Laboratory, Shanghai 200031, China\\
    \textsuperscript{\rm 3} Electrical and Computer Engineering, University of Michigan, MI 48105, United States\\
    \textsuperscript{\rm 4}School of Biomedical Engineering, Shanghai Jiao Tong University, Shanghai 200127, China\\
    \{tianxy, wuqing, zhangyy8\}@shanghaitech.edu.cn, chenlx@umich.edu, \{jiefeng, hongjiang.wei\}@sjtu.edu.cn
}

\usepackage{bibentry}

\begin{document}

\maketitle

\begin{abstract}
Cardiac magnetic resonance (CMR) imaging is widely used to characterize cardiac morphology and function. To accelerate CMR imaging, various methods have been proposed to recover high-quality spatiotemporal CMR images from highly undersampled $k$-$t$ space data. However, current CMR reconstruction techniques either fail to achieve satisfactory image quality or are restricted by the scarcity of ground truth data, leading to limited applicability in clinical scenarios. 
In this work, we proposed MoCo‑INR, a new unsupervised method that integrates implicit neural representations (INR) with the conventional motion‑compensated (MoCo) framework. Using the explicit motion modeling and the continuous prior of INRs, our MoCo-INR can produce accurate cardiac motion decomposition and high-quality CMR reconstruction. Moreover, we present a new INR network architecture tailored to the CMR problem, which can greatly stabilize model optimization.
Experiments on retrospective (\ie simulated) datasets demonstrate the superiority of MoCo‑INR over state‑of‑the‑art methods, achieving fast convergence and fine‑detailed reconstructions at ultra‑high acceleration factors (\eg 20$\times$ in VISTA sampling).
In addition, evaluations on prospective (\ie real-acquired) free‑breathing CMR scans highlight its clinical practicality for real‑time imaging. Several ablation studies also confirm the effectiveness of critical components of MoCo-INR.

\end{abstract}

\begin{links}
    \link{Code}{https://github.com/MeijiTian/MoCo-INR}
\end{links}

\section{Introduction}
Magnetic resonance (MR) imaging offers unparalleled soft tissue contrast and, as a non-invasive modality, serves as a versatile tool for evaluating cardiac function~\cite{dudley2010cardiovascular}.
However, the long acquisition time makes it difficult to capture cardiac motion accurately.
Scanning undersampled $k$-space data for each short temporal frame is an effective strategy to accelerate cardiac MR (CMR) acquisition.
Nevertheless, reconstructing artifact-free, dynamic MR images from undersampled $k$-$t$ space (\ie spatiotemporal) data poses a challenging ill-posed problem due to the violation of the Nyquist–Shannon sampling theorem~\cite{5055024}.

Many studies have proposed exploiting the inherent spatial and temporal correlations in the image sequences to alleviate the ill-posedness of the CMR problem~\cite{oscanoa2023deep}. 
A classical strategy is to incorporate a low-rank prior into the compressed sensing (CS) framework~\cite{lingala2011accelerated,zhao2011further,otazo2015low}, where the dynamic image sequence is decomposed into low-rank and sparse components. 
The integration of the low-rank prior effectively utilizes the spatiotemporal redundancy of dynamic images, thus enhancing CMR results.

In contrast, motion-compensated (MoCo) methods~\cite{pan2022learning,pan2024motion,morales201implementation} explicitly decouple frame-specific deformations from a single canonical image, allowing all temporal frames to share a common canonical spatial representation.
Thus, this decoupled modeling can achieve better CMR reconstructions for highly undersampled acquisitions.
However, most existing MoCo approaches rely on fully sampled cine CMR data for supervised training. Although effective, cine CMR acquisition requires breath-holding, which increases acquisition cost and restricts the practicality and generalizability of these methods in real-time free-breathing scenarios.

As an unsupervised learning paradigm, implicit neural representation (INR) has shown great promise in dynamic medical reconstruction~\cite{reed2021dynamic, huang2023neural,kunz2024implicit,catalan2025unsupervised,ST-INR}, where the image sequences are formulated as a continuous function of spatial–temporal coordinates. 
Thanks to the learning basis of neural networks towards low-frequency signals~\cite{xu2019frequency,rahaman2019spectral}, INR implicitly captures the spatiotemporal redundancy of dynamic images, which produces improved reconstructions. 
However, when applied to extremely ill‑posed inverse problems, the continuous prior of INR is often insufficient and requires additional regularization priors, such as image‑domain prior~\cite{kazerouni2024incode, tian2025unsupervised}, low‑rank models~\cite{ST-INR}, denoisers~\cite{iskender2025rsr}, or generative priors~\cite{du2024dper}, to enhance reconstruction quality and stability.

With the achievements of INR combined with MoCo scheme in 4D scene reconstruction~\cite{pumarola2021d,park2021nerfies}, several studies have extended this framework to 4D CT~\cite{zhang2023dynamic} and time-resolved MRI reconstruction~\cite{shao20243d,shao2025dynamic, chen2025single}. 
These approaches effectively capture respiratory motion, which is relatively simpler than cardiac motion, but often struggle to represent high‑frequency details.
Due to the high-frequency and fine detail of cardiac motion, adopting INR to achieve accurate cardiac motion decomposition from undersampled data is non-trivial.  
Meanwhile, INR is known for slow optimization, limiting its clinical practicality.
Hash‑grid encoding~\cite{muller2022instant} has been proposed to accelerate convergence and enhance high‑frequency representation.
However, its inherent discrete feature representation compromises the continuity of INR, leading to inconsistencies in continuous space and unstable optimization in dynamic reconstruction.

\par In this work, we propose \textbf{MoCo-INR}, a novel unsupervised CMR reconstruction method. Our key idea is to introduce unsupervised INRs into the MoCo framework, enabling accurate cardiac motion decomposition and the recovery of high-frequency image details. Conceptually, we explicitly decompose dynamic CMR sequences into time-varying deformations and a shared canonical image, both modeled as continuous functions parameterized by two INR networks. Benefiting from the continuous priors of INRs and the explicit motion decomposition, we effectively solve the highly ill-posed CMR inverse problem in an unsupervised manner. Moreover, we present a new INR network architecture tailored to the CMR problem, which consists of a coarse-to-fine hash encoding strategy and a CNN-based decoder. Compared to existing INR architectures, our proposed design achieves more stable optimization and produces CMR images with fine anatomical details.

We evaluate the proposed MoCo‑INR on both retrospective cine CMR reconstruction under various acquisition schemes and prospective free‑breathing CMR reconstruction.
The results demonstrate that MoCo‑INR outperforms state‑of‑the‑art (SOTA) unsupervised methods, delivering both faster convergence and higher‑fidelity reconstructions, particularly under ultra‑high acceleration factors (20$\times$ for Cartesian and 69$\times$ for non‑Cartesian).
In addition, extensive ablation studies validate the effectiveness the several key components of our MoCo-INR.
\par The main contributions as summarized as follows:
\begin{itemize}
    \item We introduce the INR to the MoCo framework, enabling accurate cardiac motion decomposition and fine-detailed reconstruction in an unsupervised manner.
    \item We propose a novel INR network architecture tailored to the CMR problem, which can greatly stabilize model optimization.
    \item We perform extensive experiments confirming the superiority of our unsupervised MoCo-INR in fast convergence and robustness with various CMR acquisitions. 
\end{itemize}

\section{Related Work}
\subsection{Motion-Compensated Approaches for CMR}

To leverage motion information, motion-compensated (MoCo) methods are introduced into CMR reconstruction to further improve performance.
MoCo methods~\cite{batchelor2005matrix,qi2021end,hammernik2021motion,camila2022self,zou2022dynamic,pan2024motion} explicitly decompose dynamic images into a canonical (or template) image and a sequence of canonical-to-observation displacement vector fields (DVFs), which can effectively exploit the spatio-temporal redundancies.
The reconstruction task is reformulated as two sub-problems: motion estimation and canonical image reconstruction, which are typically solved iteratively or within a joint optimization framework.
Thus, accurate estimation of cardiac motion is crucial to both the canonical image quality and the final reconstruction performance.
With the emergence of deep learning (DL), supervised MoCo methods~\cite{qi2021end,hammernik2021motion,pan2024motion} have gained importance in motion estimation due to their promising performance. 
However, these methods often suffer from performance degradation when the acquisition settings change (\textit{e.g.}, different sampling patterns and accelerator factors) deviate from the training data.
Moreover, the long acquisition time of MRI makes it difficult to obtain high-quality ground-truth labels.
These issues pose substantial obstacles to the practical application of supervised methods in clinical settings.
Current unsupervised MoCo methods have mainly focused on respiratory motion~\cite{camila2022self,zou2022dynamic} but have rarely explored cardiac motion~\cite{Kettelkamp2023Motion}, which is more complex and requires higher temporal resolution.

\subsection{INR for Dynamic MRI Reconstruction}
Recently, many INR-based dynamic MRI reconstruction methods~\cite{huang2023neural, catalan2025unsupervised, ST-INR} have been proposed, formulating dynamic MRI image sequences as spatial–temporal functions represented in either the image domain or in $k$-space. 
While INR effectively exploits spatial–temporal correlations to constrain the reconstruction, existing methods are known for their slow convergence, often requiring hours to reconstruct a single slice~\cite{kunz2024implicit}, particularly when modeling high‑frequency features.
The current SOTA approach~\cite{ST-INR} adopts hash‑grid encoding to accelerate convergence but still relies on additional low‑rank and sparsity constraints to ensure consistent reconstruction quality.
However, explicit motion‑compensated representations remain unexplored in improving the robustness and efficiency of INR‑based optimization in real-time cardiac MRI reconstruction.

\begin{figure*}[!t]
    \includegraphics[width=\textwidth]{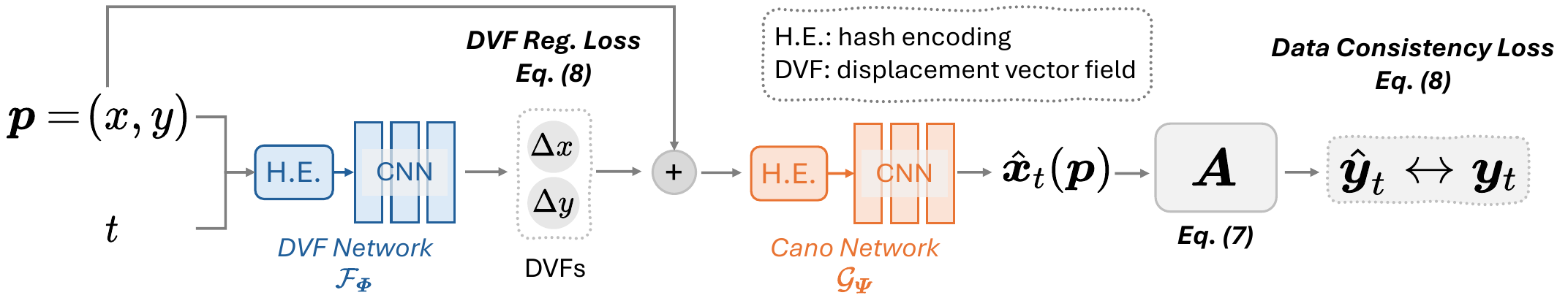}
    \caption{Overview of the proposed MoCo-INR framework. Given any spatial coordinate $\boldsymbol{p}=(x,y)$ in the physical space and temporal coordinate $t$,  our deformation network predicts the corresponding time-varying displacement vector field (DVF).
    Adding this displacement to the spatial coordinate in physical space yields the associated location in the canonical space. 
    Then, the canonical network maps these warped coordinates to the dynamic image $\boldsymbol{x}_t$. Finally, the two networks are jointly optimized by minimizing the data-consistency loss (Eq.~\ref{eq:Loss}) and DVF regularization loss (Eq.~\ref{eq:Loss}).  
}
\label{method:pipeline}
\end{figure*}

\section{Preliminaries}
\subsection{Forward Model of Dynamic CMR}
The forward physical model of dynamic CMR acquisition can be expressed as:
\begin{equation}
\boldsymbol{y}_{t, c} = \boldsymbol{M}_{t}\mathcal{T}\boldsymbol{S}_{c}\boldsymbol{x}_t + \boldsymbol{n}_{{t}, c},
\label{eq:forward_model}
\end{equation}
where $\boldsymbol{x}_t \in \mathbb{C}^{N}$ is the image at any timestemp $t=1,\dots,T$ and $\boldsymbol{S}_{c} \in \mathbb{C}^{N\times N}$ represents the $c^\text{th}$ coil sensitivity map. $\mathcal{T}$ is the Fourier transform operator, $\boldsymbol{M}_{t} \in \mathbb{R}^{M\times N}$ is the binary diagonal undersampling matrix, $\boldsymbol{n}_{{t}, c} \in \mathbb{C}^{M}$ is the system noise assuming Gaussian distribution, and $\boldsymbol{y}_{t, c} \in \mathbb{C}^{M}$ is the acquired $k$-space measurement. 

\subsection{Motion-Compensated Representation}
To fully exploit the spatiotemporal redundancy in the image sequence $\{\boldsymbol{x}_t\}_{t=1}^T$ and alleviate the ill-posedness of the CMR inverse problem, motion-compensated (MoCo) representation decouples each frame $\boldsymbol{x}_t$ into a shared canonical image $\boldsymbol{x}_\text{cano}$ and a corresponding displacement vector field (DVF) $\boldsymbol{u}_t$. Formally, this can be expressed as:
\begin{equation}
    \boldsymbol{x}_t = \mathcal{W}(\boldsymbol{x}_\text{cano}, \boldsymbol{u}_t),
\end{equation}
where $\mathcal{W}$ denotes the image warping operator. The DVF defines, for each voxel of the frame $\boldsymbol{x}_t$, the offset $(\Delta x, \Delta y)_t$ between the canonical space and its physical location. \par Conventional MoCo approaches for dynamic MRI represent the canonical image $\boldsymbol{x}_\text{cano}$ as a discrete matrix. 
The warping operator is then used with the DVFs $\{\boldsymbol{u}_t\}_{t=1}^T$ to interpolate this matrix and generate the image sequence $\{\boldsymbol{x}_t\}_{t=1}^T$.
Although effective, discrete interpolation may lose high-frequency details and thus limit reconstruction quality.

\section{Proposed Method}
Our goal is to reconstruct artifact-free CMR images with high spatiotemporal resolution from highly undersampled $k$-$t$ space data in \textit{an unsupervised way}.
To this end, we propose MoCo-INR, a new unsupervised CMR method that first integrates INR into the MoCo framework.
Thanks to the continuous representation enabled by INR, MoCo-INR can achieve accurate estimation of cardiac motion and image reconstructions with preserved high-frequency details.

\subsection{Continuous Representations of DVFs and Canonical Image}
To accurately recover both DVFs $\{\boldsymbol{u}_t\}_{t=1}^T$ and the shared canonical image $\boldsymbol{x}_{\text{cano}}$, we leverage INR to formulate them in a continuous form, instead of discrete matrices as in traditional MoCo-based methods. Specifically, the DVFs are defined as a single continuous function $\boldsymbol{f}$ of spatial-temporal coordinate, as below:
\begin{equation}
 \boldsymbol{f}:\quad(\boldsymbol{p}, t) \in \mathbb{R}^3 \mapsto \boldsymbol{u}_{t} (\boldsymbol{p}) = (\Delta x, \Delta y) \in \mathbb{R}^2,
\end{equation}
where $(\boldsymbol{p}, t)$ denotes any spatial-temporal coordinate in the physical space, and $\boldsymbol{u}_t(\boldsymbol{p})$ is the displacement vector to map $\boldsymbol{p}$ into the canonical space. While the complex-valued canonical image $\boldsymbol{x}_\text{cano}$ is formulated as a continuous function $\boldsymbol{g}$ of spatial coordinate, as below:
\begin{equation}
    \boldsymbol{g}:\quad\tilde{\boldsymbol{p}} \in \mathbb{R}^{2} \mapsto \boldsymbol{x}_\text{cano}(\tilde{\boldsymbol{p}})=a(\tilde{\boldsymbol{p}})+jb(\tilde{\boldsymbol{p}}) \in \mathbb{C},
\end{equation}
where $\tilde{\boldsymbol{p}}$ represents any coordinate in a 2D canonical space, and $\boldsymbol{x}_\text{cano}(\tilde{\boldsymbol{p}})$ is the corresponding complex‑valued intensity. 
\par MoCo-INR uses a DVF network $\mathcal{F}_{\boldsymbol{\varPhi}}$ and a canonical network $\mathcal{G}_{\boldsymbol{\varPsi}}$ to approximate the two functions, respectively. Technically, $\mathcal{F}_{\boldsymbol{\varPhi}}$ takes spatial-temporal coordinates as input and outputs the DVF estimations (\ie $\boldsymbol{u}_{t}(\boldsymbol{p}) = \mathcal{F}_{\boldsymbol{\varPhi}}(\boldsymbol{p}, t)$), while $\mathcal{G}_{\boldsymbol{\varPsi}}$ takes spatial coordinates as input and predicts the real and imaginary parts of the canonical image (\ie $[a(\tilde{\boldsymbol{p}}), b(\tilde{\boldsymbol{p}})] = \mathcal{G}_{\boldsymbol{\varPsi}}(\tilde{\boldsymbol{p}})$). Benefiting from the learning bias toward low-frequency continuous signals~\cite{xu2019frequency,rahaman2019spectral}, the continuous functions $\boldsymbol{f}$ and $\boldsymbol{g}$ can be well approximated, enabling high-quality reconstructions of both the DVFs and canonical image.

\subsection{Model Optimization}
Fig.~\ref{method:pipeline} shows the workflow of MoCo-INR, where we jointly optimize the DVF $\mathcal{F}_{\boldsymbol{\varPhi}}$ and the canonical network $\mathcal{G}_{\boldsymbol{\varPsi}}$.
\paragraph{Prediction of CMR Image.}
\par Given the acquired $k$-space data $\boldsymbol{y}_t$ at any timestemp $t=1,\dots,T$, we first feed the spatial-temporal coordinate $(\boldsymbol{p},t)$, defined in the physical space, into the network $\mathcal{F}_{\boldsymbol{\varPhi}}$ to predict the corresponding DVF $\boldsymbol{u}_t(\boldsymbol{p})$. This DVF is then used to transform the spatial coordinate from the physical space to the canonical space. Formally, this process can be expressed as:
\begin{equation}
    \begin{aligned}
        \tilde{\boldsymbol{p}} =\boldsymbol{p} + \boldsymbol{u}_t(\boldsymbol{p}),\quad\text{with}\quad\boldsymbol{u}_t(\boldsymbol{p})=\mathcal{F}_{\boldsymbol{\varPhi}}(\boldsymbol{p}, t).
    \end{aligned}
\end{equation}
\par Then, the canonical network $\mathcal{G}_{\boldsymbol{\varPsi}}$ takes the warped coordinates $\tilde{\boldsymbol{p}}$ as input and estimates the corresponding dynamic image as follows:
\begin{equation}
    \hat{\boldsymbol{x}}_t(\tilde{\boldsymbol{p}}) = \mathcal{G}_{\boldsymbol{\varPsi}}(\tilde{\boldsymbol{p}}).
\end{equation}
\paragraph{Differentiable Forward Model.}
\par According to the forward acquisition of dynamic CMR (Eq.~\ref{eq:forward_model}), we generate the $k$-space data estimations $\hat{\boldsymbol{y}}_t$ from the predicted CMR image $\hat{\boldsymbol{x}}_t$, which is defined as follows:
\begin{equation}
    \boldsymbol{A}:\quad\hat{\boldsymbol{y}}_t = \boldsymbol{M}_t\mathcal{T}\boldsymbol{S}_c\hat{\boldsymbol{x}}_t,
\end{equation}
where the operators $\boldsymbol{M}_t$, $\mathcal{T}$, and $\boldsymbol{S}_c$ depend on the CMR acquisition protocols and are known. More importantly, the forward model $\boldsymbol{A}$ is differentiable, allowing the use of gradient descent-based backpropagation algorithms (\eg Adam) for network optimization.
\subsubsection{Loss Function.}
Finally, the DVF network $\mathcal{F}_{\boldsymbol{\varPhi}}$ and the canonical network $\mathcal{G}_{\boldsymbol{\varPsi}}$ are jointly optimized by minimizing the following loss function $\mathcal{L}$ as below:
\begin{equation}
    \begin{aligned}
            &\mathcal{L} = \underbrace{\big\|\hat{\boldsymbol{y}}_t-\boldsymbol{y}_t\big\|_1}_{\mathcal{L}_{\text{DC}}} +\ \mathcal{L}_\text{DVF},\\
        &\text{with}\quad\mathcal{L}_\text{DVF} =\big\| \boldsymbol{u}_t \big\|_1 + \big\|\nabla \boldsymbol{u}_t\big\|_1 +  \big\|\nabla^2 \boldsymbol{u}_t\big\|_1,
    \end{aligned}
\label{eq:Loss}
\end{equation}
where $\mathcal{L}_\text{DC}$ represents the data consistency term that minimizes the distance between the acquired and estimated $k$-space data. $\mathcal{L}_\text{DVF}$ is a sparsity and smoothness regularization term that enforces plausible DVF estimations and further stabilizes the joint network optimization. Its effectiveness is explored in the following experiments.
\subsubsection{CMR Image Reconstruction.}
\par After model optimization, the high-quality CMR sequence $\{\boldsymbol{x}_t^*\}_{t=1}^T$ can be directly reconstructed by feeding all spatial-temporal coordinates $(\boldsymbol{p}, t)$ into the well-trained MoCo-INR model.

\subsection{INR Network Designed for CMR Problem}
\par Traditional INR networks often consist of a coordinate encoding module, such as Hash encoding~\cite{muller2022instant}, and an MLP decoder. These encoding modules can greatly enhance the network's ability to capture high-frequency signals, improving detail recovery. However, applying existing INR networks in CMR often yields unsatisfactory performance due to the problem’s ill-posedness and strong nonlinearity. To address this, we propose a novel coarse-to-fine hash encoding and a CNN-based decoder to achieve reliable DVF estimation and detailed image reconstruction.

\subsubsection{Coarse-to-Fine Hash Encoding.}
\begin{figure}[!t]
    \centering
    \includegraphics[width=\linewidth]{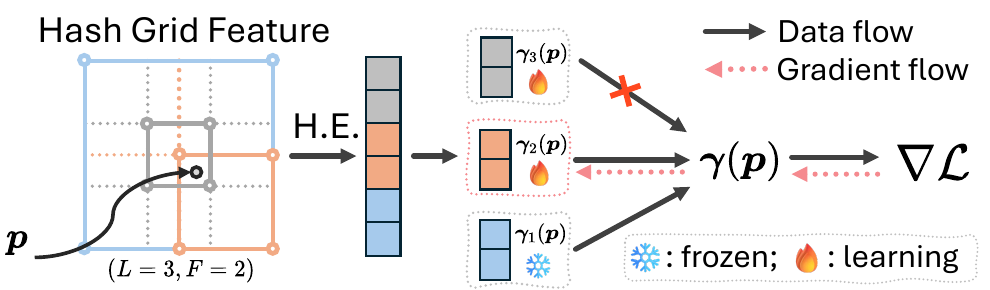}
    \caption{Illustration of the proposed coarse-to-fine hash encoding strategy. Given any input coordinate $\boldsymbol{p}$, the low-frequency feature (\ie $\boldsymbol{\gamma}_1$) is learned first and then frozen. As the optimization proceeds, higher-frequency features (\ie $\boldsymbol{\gamma}_2$ and $\boldsymbol{\gamma}_3$) are progressively optimized.}
    \label{fig:c2f_illustration}
\end{figure}

Hash encoding~\cite{muller2022instant} is a cutting-edge encoding strategy. It maps low-dimensional coordinates $\boldsymbol{p}$ into high-dimensional features $\boldsymbol{\gamma}(\boldsymbol{p}) = \boldsymbol{\gamma}_1(\boldsymbol{p}) \oplus \cdots \oplus\boldsymbol{\gamma}_l(\boldsymbol{p}) \oplus \cdots \oplus \boldsymbol{\gamma}_L(\boldsymbol{p}) \in \mathbb{R}^{LF}$, where each $\boldsymbol{\gamma}_l(\boldsymbol{p}) \in \mathbb{R}^F$ denotes an $F$-dimensional feature at the $l$-th resolution level. The low-resolution features (\eg $\boldsymbol{\gamma}_1$) capture low-frequency global structures, while the high-resolution ones (\eg $\boldsymbol{\gamma}_L$) model high-frequency local details. A recent study on MRI reconstruction~\cite{wu2025moner} demonstrated that the global structures are more crucial for rigid motion correction. Inspired by this observation, we propose a novel coarse-to-fine scheme for the CMR problem. As shown in Fig.~\ref{fig:c2f_illustration}, the optimization starts by learning the low-frequency features to capture global motion. Then, the higher-frequency features are progressively optimized to refine fine-scale motion details. This coarse-to-fine hash encoding can enhance reliable DVF estimations, thus enabling improved CMR reconstructions.

\subsubsection{CNN-based Decoder.}
\par Existing INR networks typically use an MLP as decoder to transform the encoded features into target signals. Although effective, the voxel-wise mapping of MLP-based decoders struggles to capture the spatial continuity of images~\cite{mihajlovic2024resfields}. Moreover, the powerful fitting capability introduced by the encoding may further lead to overfitting to undersampled data, resulting in high-frequency artifacts. To address these issues, we introduce a three-layer convolutional neural network (CNN) to replace the conventional MLP decoder. Owing to the inductive bias of CNNs toward local structures, the continuous functions $\boldsymbol{f}$ and $\boldsymbol{g}$, which represent the DVFs and the canonical image, can be better approximated, thereby improving the quality of reconstructed CMR images.

\begin{figure*}[t]
\centering
\includegraphics[width=\textwidth]{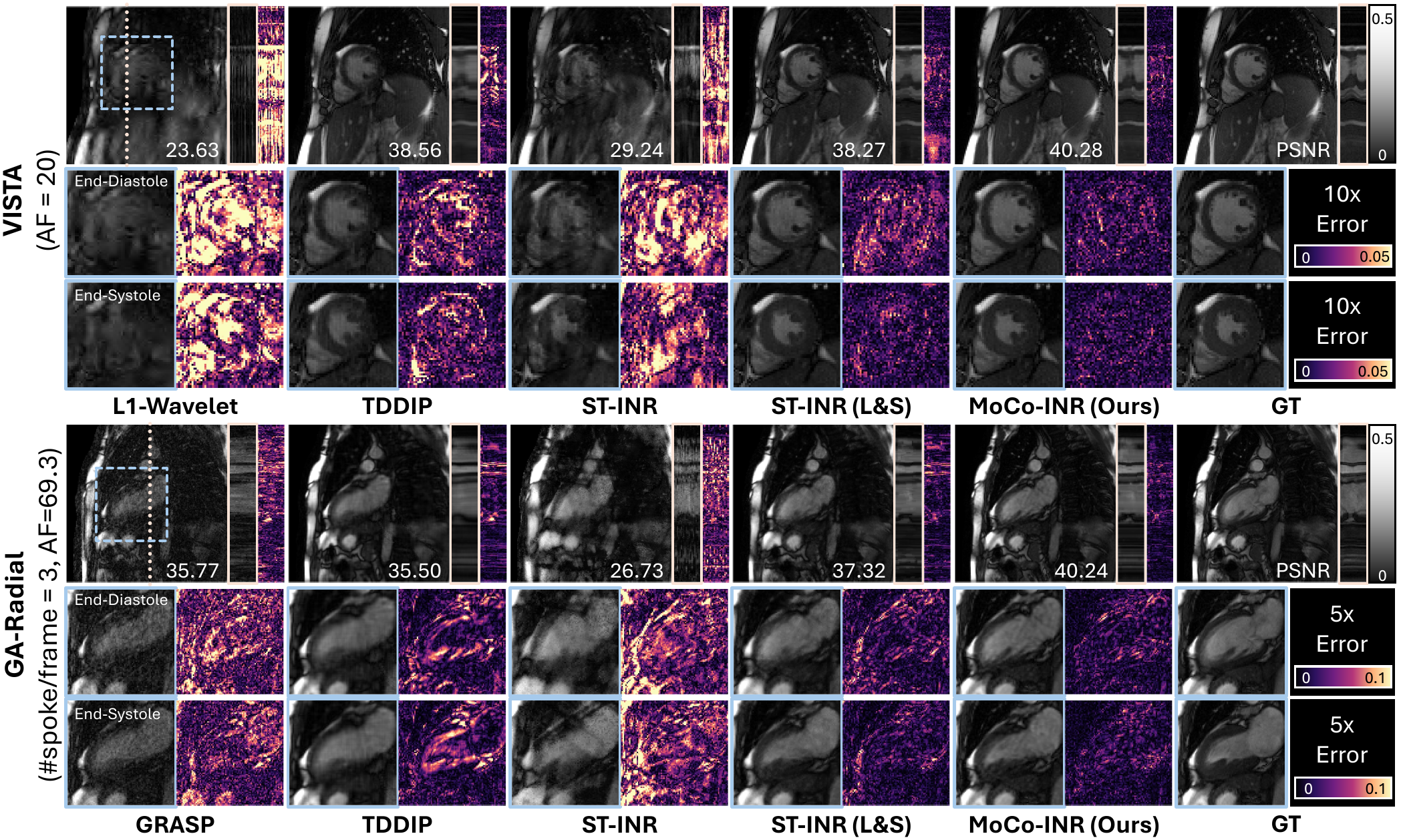} 
\caption{
Qualitative results of retrospective reconstructions obtained with the compared methods. The figure displays the reconstructed image, its profile line over time (the $y\text{-}t$ plane), and the corresponding error map. The selected $y$-axis is indicated by a white dashed line, and zoom-in boxes highlight regions of interest at the end-diastole (ED) and end-systole (ES) phases.
The upper part shows results for SAX slice acquired using a VISTA sampling pattern with an acceleration factor of AF=20. The bottom part shows results for LAX slice acquired using a golden-angle radial sampling pattern with 3 spokes.
}
\label{res_fig1}
\end{figure*}

\section{Experimental Settings}

\subsection{Retrospective Reconstruction Study} 
\subsubsection{Dataset.}
We used the fully sampled cardiac cine dataset from the public OCMR dataset~\cite{chen2020ocmr}.
All scans were acquired using prospective ECG gating and breath-holding.
For this study, we selected 11 slices, including five long-axis (LAX) views and six short-axis (SAX) views, acquired on a 1.5T clinical MRI scanner (Magnetom Aera, Siemens Healthineers).
\subsubsection{Simulation Process.}
The original data were cropped into a square shape and resized to a size of  $208 \times 208$.
To evaluate reconstruction performance under different sampling strategies, we simulated both Cartesian and non-Cartesian undersampling.
\textbf{Cartesian}: we adopted VISTA pattern with two accerelation factor (AF) of 12 and 20. 
\textbf{Non-Cartesian}: we adopted golden-angle (GA) radial pattern combined with the NUFFT operator~\cite{fessler2003nonuniform}, using 8 and 3 spokes per frame, corresponding to AF of 26 and 69.3.

\subsection{Prospective Reconstruction Study}
We used prospectively acquired real-time CMR data from the public OCMR dataset~\cite{chen2020ocmr}, under \textbf{free-breathing} conditions with VISTA sampling mask and the acceleration factor of AF=9.
Ten slices of SAX view were selected for the study, each with an in-plane resolution of $2.08\times 2.08 \text{mm}^2$ and a slice thickness of 8mm. 
Each slice comprises 65 temporal frames, corresponding to a temporal resolution of 38.4 ms ($\approx$26 Hz).

\begin{table*}[!t]
\setlength{\tabcolsep}{1.4mm}
\centering
\begin{tabular}{cccccccc} 
\toprule  
\textbf{Sampling}          & \textbf{AF}                & \textbf{Metric} & \textbf{$\ell_1$-Wavelet} & \textbf{TDDIP} & \textbf{ST-INR} &\textbf{ST-INR (L\&S)} & \textbf{MoCo-INR}  \\
\midrule
\multirow{6.5}{*}{VISTA} & \multirow{3}{*}{12$\times$} & PSNR    &  28.00$\pm$1.98$^{***}$     &   38.05$\pm$2.99$^{***}$    &  36.31$\pm$2.62$^{***}$        &   \underline{41.35$\pm$2.60}$^{***}$            &  \textbf{42.25$\pm$2.64}     \\
                  &                   & SSIM    &   0.734$\pm$0.038$^{***}$    &   0.943$\pm$0.025$^{***}$    &   0.934$\pm$0.020$^{***}$       &     \textbf{0.972$\pm$0.012}$^\blacktriangledown$           & \underline{0.971$\pm$0.013}      \\

                  &                   & nRMSE (ROI)   &  0.450$\pm$0.088$^{***}$      &   0.206$\pm$0.037$^{***}$    &    0.150$\pm$0.022$^{***}$     &   \underline{0.109$\pm$0.024}$^{***}$             &  \textbf{0.093$\pm$0.017}     \\
                  
\cmidrule(){2-8}
  & \multirow{3}{*}{20$\times$} & PSNR   &  23.82$\pm$1.69$^{***}$     & \underline{36.58$\pm$2.70}$^{***}$      &   31.24$\pm$3.00$^{***}$       & {36.26$\pm$2.94}$^{***}$        &  \textbf{39.53$\pm$2.58}       \\
                  &                   & SSIM   & 0.576$\pm$0.039$^{***}$      &   0.929$\pm$0.026$^{***}$    &  0.843$\pm$0.042$^{***}$        &     \underline{0.937$\pm$0.021}$^{***}$          & \textbf{0.957$\pm$0.017}       \\
                                    &                   & nRMSE (ROI)  &    0.658$\pm$0.044$^{***}$   &  0.217$\pm$0.037$^{***}$     &         0.229$\pm$0.030$^{***}$&  \underline{0.158$\pm$0.024}$^{***}$              &    \textbf{0.125$\pm$0.022}   \\
\midrule
\midrule
\textbf{Sampling}          & \textbf{AF}                & \textbf{Metric} & \textbf{GRASP} & \textbf{TDDIP} & \textbf{ST-INR} &\textbf{ST-INR (L\&S)} & \textbf{MoCo-INR}  \\
\midrule
\multirow{6.5}{*}{\begin{tabular}[c]{@{}c@{}}GA\\Radial\end{tabular}} & \multirow{3}{*}{26.0$\times$} & PSNR   &  32.14$\pm$3.33$^{***}$     &   34.10$\pm$1.97$^{***}$    &  30.96$\pm$2.04$^{***}$        &   \underline{38.85$\pm$2.86}$^{**}$            &  \textbf{40.33$\pm$2.48}     \\
                  &                   & SSIM   &   0.886$\pm$0.037$^{***}$    &   0.895$\pm$0.022$^{***}$    &   0.812$\pm$0.028$^{***}$       &     \underline{0.956$\pm$0.016}$^{*}$           & \textbf{0.960$\pm$0.016}      \\
                                    &                   & nRMSE (ROI)  &  0.253$\pm$0.056$^{***}$     & 0.227$\pm$0.033$^{***}$      & 0.219$\pm$0.040$^{***}$        &      \underline{0.118$\pm$0.014}$^{*}$          &   \textbf{0.109$\pm$0.012}    \\
\cmidrule(){2-8}
  & \multirow{3}{*}{69.3$\times$} & PSNR   &  26.24$\pm$2.24$^{***}$     & 33.62$\pm$2.10$^{***}$      &   26.58$\pm$1.82$^{***}$       & \underline{33.92$\pm$3.13}$^{***}$        &  \textbf{37.75$\pm$2.53}       \\
                  &                   & SSIM   & 0.717$\pm$0.038$^{***}$      &   0.883$\pm$0.028$^{***}$    &  0.682$\pm$0.039$^{***}$        &     \underline{0.910$\pm$0.034}$^{**}$          & \textbf{0.940$\pm$0.024}       \\
                                    &                   & nRMSE (ROI)  &  0.422$\pm$0.118$^{***}$     &0.238$\pm$0.034$^{***}$       &   0.338$\pm$0.084$^{***}$      &        \underline{0.196$\pm$0.040}$^{*}$        &    \textbf{0.165$\pm$0.022}   \\
\bottomrule
\end{tabular}
        \begin{tablenotes}[flushleft]
            \footnotesize
            \item  The best and second performances are highlighted in \textbf{bold} and \underline{underline}. Statistical significant differences compared with our MoCo-INR are marked ($^{***}$ $p<$0.001; $^{**}$ $p<$0.01; $^{*}$ $p<$0.05; and $^\blacktriangledown$ $p\ge$0.05, not significant).
        \end{tablenotes}
\caption{Quantitative results (PSNR (dB)/SSIM/nRMSE) of the compared methods under a Cartesian sampling pattern (VISTA) and a non-Cartesian sampling pattern (GA Radial) under two acceleration factors, respectively. 
}
\label{tab1}
\end{table*}


\subsection{Methods in Comparison \& Metrics }
\subsubsection{Methods in Comparison.}
We compare our method with five representative unsupervised methods: (1) Compressed-sensing (CS) based: \textbf{$\ell_1$-Wavelet}~\cite{l1wavelet} for Cartesian sampling; \textbf{GRASP}~\cite{GRASP} for golden-angle radial sampling; 
(2) DIP-based: \textbf{Time-Depend DIP (TDDIP)}~\cite{TDDIP}; (3) INR-based: \citet{ST-INR} proposed an INR‑based dynamic MRI reconstruction method that incorporates hash encoding with additional low‑rank and sparsity (L\&S) constraints; we refer to this approach as \textbf{ST-INR (L\&S)}. To evaluate the effectiveness of these additional constraints, we also include a variant without the (L\&S) constraints, denoted simply as \textbf{ST‑INR}.

\subsubsection{Evaluation Metrics.}
For the reconstructed CMR images, we employ peak signal-to-noise ratio (PSNR) and structural similarity index (SSIM) as quantitative evaluation metrics. 
To specifically evaluate the accurate reconstruction of cardiac anatomy and its temporal dynamics, we segment the cardiac region and compute the normalized root-mean-square error (nRMSE) within it.
To quantify reconstruction efficiency, we present the runtime for DL-based methods to achieve the reported optimal performance.

\begin{figure}[!t]
\centering
\includegraphics[width=0.95\linewidth]{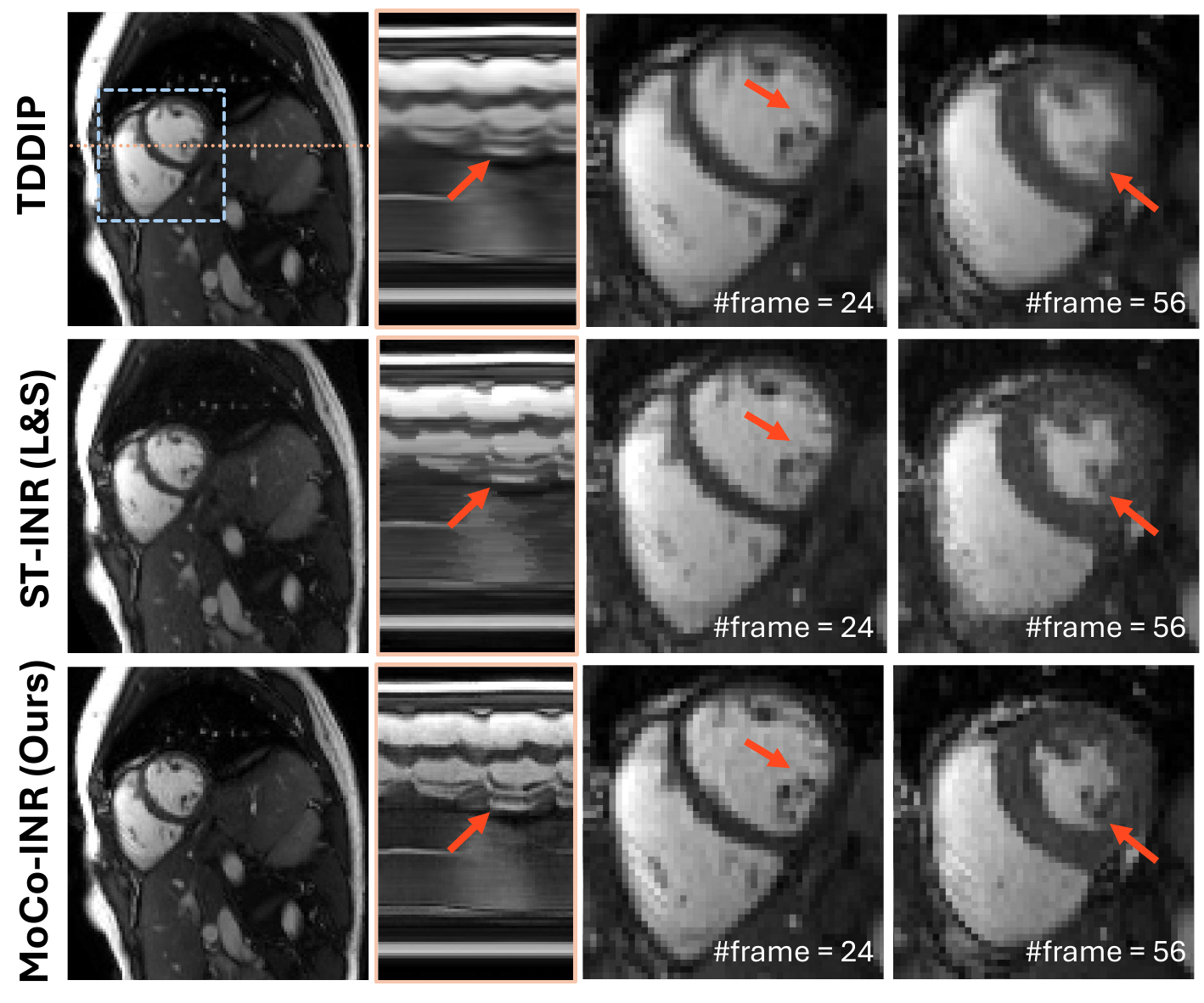} 
\caption{Qualitative results of prospective reconstruction under free-breathing scans.
}
\label{res_fig2}
\end{figure}

\subsection{Implementation Details}
In MoCo-INR, the DVF network $\mathcal{F}_{\boldsymbol{\varPhi}}$ adopts hash encoding set of $N_\text{min}=2$, $T=2^{21}$, $L=10$, $F=4$ and $b=2$, while the canonical network $\mathcal{G}_{\boldsymbol{\varPsi}}$ is set as follows $N_\text{min}=2$, $T=2^{21}$, $L=12$, $F=8$ and $b=2$. 
Both networks employ lightweight CNN decoders composed of three convolutional layers. The first two convolutional layers are followed by nonlinear activation functions, with $64$ filters of size of $3$, and the final convolutional layer outputs without activation.
Due to the space constraint, we introduce the other implemental details of MoCo-INR and baselines in the supplemental materials.

\section{Results}
\subsection{Retrospective Reconstruction Results}
\begin{figure}[!t]
\centering
\includegraphics[width=0.95\linewidth]{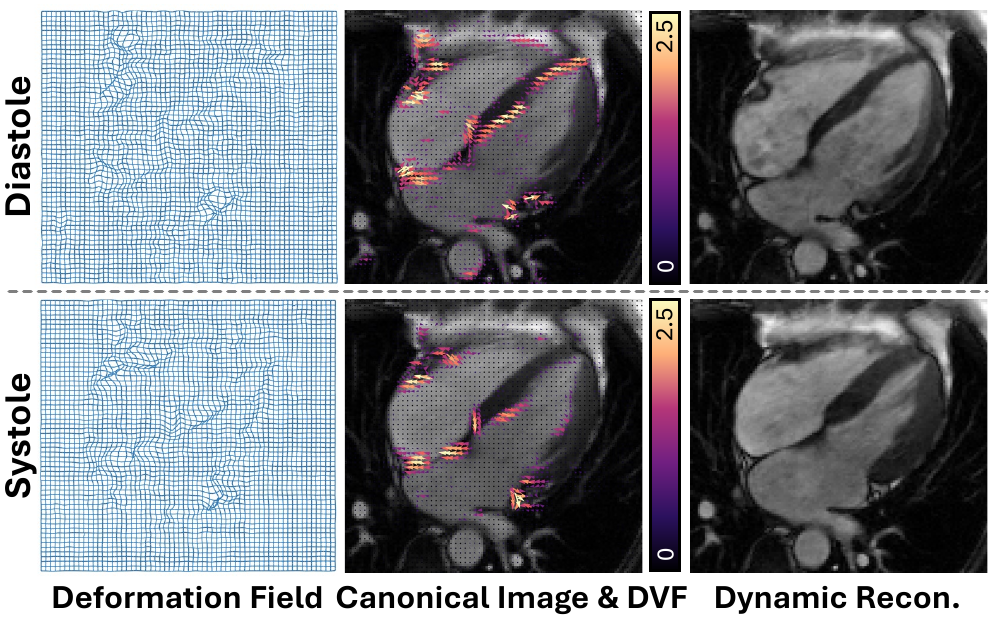} 
\caption{Visualization of the estimated DVFs and canonical image of MoCo-INR at the diastolic and systolic phases. 
}
\label{res_dvf}
\end{figure}

Table~\ref{tab1} compares the performance of our MoCo-INR with baselines. 
Under ultra-high acceleration factors (20$\times$ for Cartesian and 69$\times$ for non‑Cartesian), MoCo-INR attains the highest PSNR/SSIM values, with improvements that are statistically significant ($p<$0.001). 
The results highlight its robustness to severe undersampling and demonstrate its suitability for challenging reconstruction scenarios. 
Moreover, MoCo-INR consistently yields the lowest ROI nRMSE across every sampling pattern and acceleration setting, confirming its superior ability to preserve both the dynamic motion and the anatomical detail of the cardiac region.

Fig.~\ref{res_fig1} shows the qualitative results of reconstruction.
The $\ell_{1}$-Wavelet method fails to recover images at high acceleration factors, exhibiting severe blurring and aliasing artifacts.
TDDIP insufficiently captures temporal dynamics. 
Particularly, under golden-angle radial sampling, the cardiac anatomy appears nearly identical between ED and ES phases.
ST-INR introduces numerous artifacts due to its lack of explicit regularization, while ST-INR (L\&S) mitigates these artifacts yet still suffers from noticeable noise and indistinct tissue boundaries.
In contrast, the proposed MoCo-INR exploits shared spatial information across temporal frames within a motion-compensation framework and employs a CNN decoder that robustly processes hash-grid features, thereby enabling accurate cardiac motion tracking and high-fidelity anatomical reconstruction.

\subsection{Prospective Reconstruction Results}

\begin{table}[!t]
\setlength{\tabcolsep}{1mm}
\centering
\begin{tabular}{ccccc} 
\toprule
\textbf{Study}                  & \textbf{Sampling}  & \textbf{TDDIP} & \textbf{ST-INR (L\&S)} & \textbf{MoCo-INR}  \\
\midrule
\multirow{2}{*}{\textbf{Retro.}} & VISTA     &  3.2     &  \underline{1.5}       & \textbf{1.3}          \\

                        & GA Radial &  23.3     &  \underline{10.9}      & \textbf{5.5}          \\
\midrule
\textbf{Prosp.}                 & VISTA     &   19.3    &  \underline{6.7}       &   \textbf{3.4}        \\
\bottomrule
\end{tabular}
\caption{Runtime (in minutes) comparisons for the unsupervised DL-based methods.}
\label{tab2}
\end{table}
We further evaluate the proposed MoCo-INR and the compared methods on prospectively undersampled data. 
The visual comparison is illustrated in Fig.~\ref{res_fig2}, consistent with the observations from the retrospective study.
TDDIP exhibits over-smoothing in both spatial and temporal dimensions, and the zoom-in views reveal anatomically implausible structures.
Compared with ST-INR (L\&S), the proposed MoCo-INR yields sharper tissue detail with significantly reduced artifacts, as highlighted by the red arrows.
Notably, the intensity profile shows that, although ST-INR (L\&S) suffers from spatial noise, it fails to capture temporal detail, whereas MoCo-INR successfully resolves both large-scale cardiac motion and subtle intramural deformations.

\subsection{Evaluation of DVFs and Canonical Image}
Fig.~\ref{res_dvf} shows the estimated DVFs alongside the canonical image. 
The quiver plots (second column) illustrate vector patterns consistent with myocardial relaxation and chamber enlargement during diastole, and with myocardial contraction and ventricular volume reduction during systole.
The learned DVFs are consistent with the known biomechanics of the cardiac cycle, demonstrating our method accurately capture cardiac motion.

\subsection{Evaluation of Runtime}
Fast runtime is essential for clinical applicability.
Table~\ref{tab2} reports the runtime on a single NVIDIA RTX 4090 GPU, showing that the proposed MoCo‑INR achieves fast reconstruction for both retrospective and prospective studies. 
By explicitly decomposing deformation and canonical image content, MoCo‑INR enables faster convergence with fewer optimization steps compared to ST‑INR (L\&S).
This significantly reduces the computational cost, particularly for non‑Cartesian sampling where the NUFFT operator is inherently slow.

\subsection{Ablation Studies}
\subsubsection{Effectiveness of CNN-based INR network.}

\begin{figure}[!t]
\centering
\includegraphics[width=0.95\linewidth]{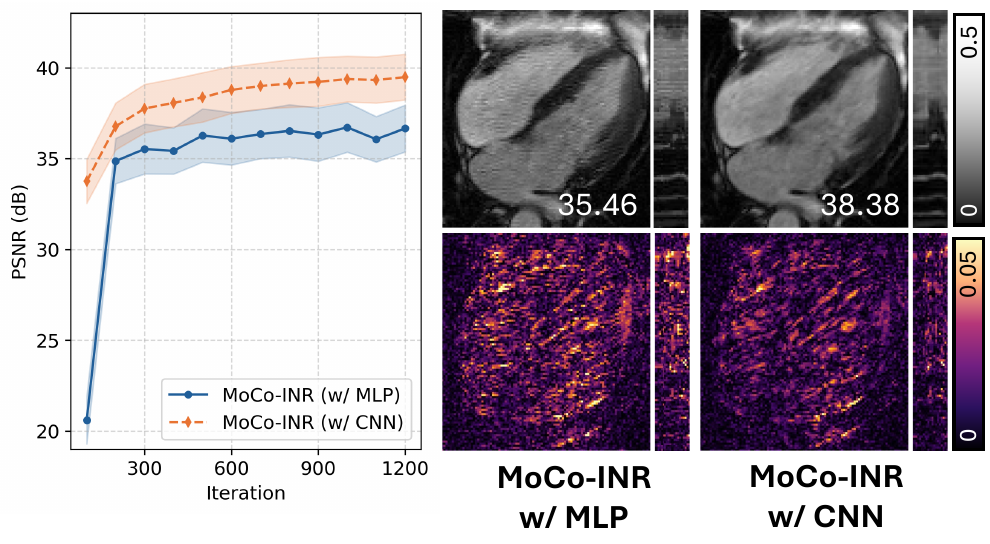} 
\caption{Performance curves and qualitative results for VISTA sampling with AF=20 of MoCo-INR with either an MLP decoder or a CNN decoder.
}
\label{abl_dec}
\end{figure}

Fig.~\ref{abl_dec} compares the performance of MoCo‑INR using an MLP decoder versus a CNN decoder.
The performance curves show that the CNN‑based decoder consistently outperforms the MLP decoder and provides a more stable optimization process.
In the reconstructed MR images, the MLP decoder introduces noticeable high‑frequency artifacts, whereas the CNN decoder produces smoother and more accurate results, as further highlighted in the error maps.

\subsubsection{Influence of DVF Regularization and Coarse2fine Hash Encoding Learning.}
\begin{table}[!t]
\centering
\begin{tabular}{ccc} 
\toprule
\textbf{Model} & \textbf{PSNR} & \textbf{SSIM}  \\
\midrule
w/o $\mathcal{L}_\text{DVF}$      &  34.42$\pm$2.73$^{***}$    & 0.895$\pm$0.031$^{***}$      \\
w/o Coarse2fine      &  35.51$\pm$2.84$^{***}$    &  0.926$\pm$0.026$^{***}$     \\
Full      & \textbf{37.75$\pm$2.53}     & \textbf{0.940$\pm$0.024}      \\
\bottomrule
\end{tabular}
\caption{Quantitative comparisons on retrospective study using MoCo-INR, evaluated without key components under golden-angle radial sampling with AF=69.3. }
\label{tab3}
\end{table}
\begin{figure}
    \centering
    \includegraphics[width=0.95\linewidth]{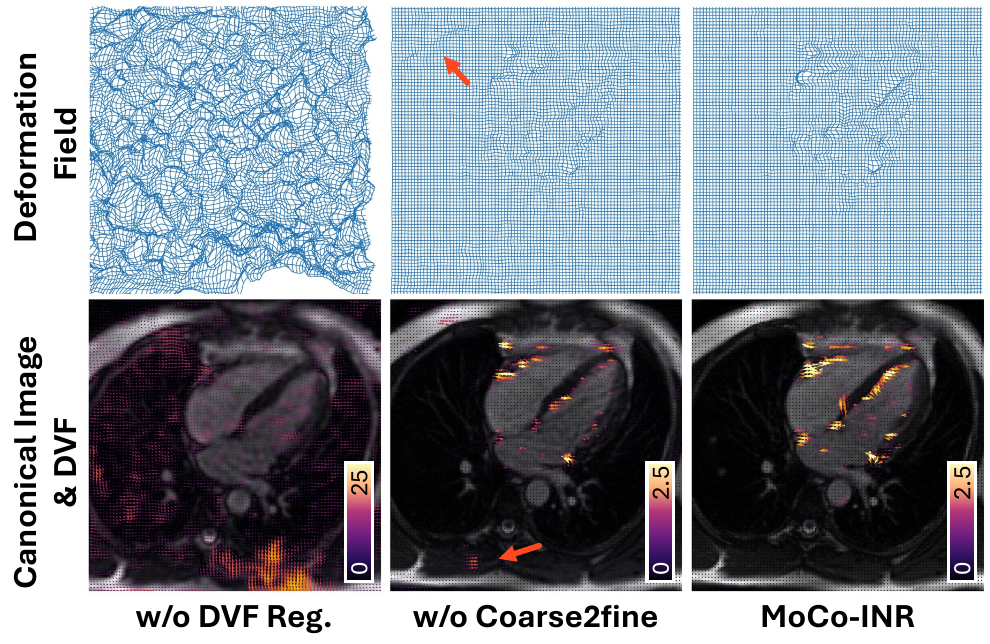}
    \caption{Qualitative comparison showing the influence of DVF regularization and the coarse‑to‑fine hash‑encoding learning strategy of MoCo‑INR on DVF estimation and canonical image reconstruction. }
    \label{fig:dvf_abl}
\end{figure}

Table~\ref{tab3} demonstrates the effectiveness of DVF regularization and the coarse‑to‑fine learning strategy, showing a significant degradation in reconstruction performance when these components are removed.
Fig.~\ref{fig:dvf_abl} further illustrates their influence.
Without DVF regularization, the estimated DVF is largely incorrect.
When the coarse‑to‑fine learning strategy is not applied, the DVF estimation is relatively reasonable but still exhibits abnormal motion in static regions (highlighted by orange arrows).
In contrast, MoCo‑INR with the proposed coarse‑to‑fine learning accurately captures plausible cardiac motion.

\section{Conclusion \& Discussion}

This work introduces MoCo-INR, a novel unsupervised motion-compensated framework for cardiac MR reconstruction.
Experimental results show that MoCo-INR achieves superior performance under ultra-high acceleration factors acquisitions and is capable of accurately reconstructing real‑time free‑breathing scans.
Benefiting from the flexibility of unsupervised nature and fast convergence, MoCo-INR is well‑suited for a variety of acquisition conditions encountered in clinical practice. 
Despite these promising results, several challenges remain. 
Future work will focus on extending MoCo‑INR to high‑resolution 3D spatial–temporal reconstructions and addressing limitations of motion compensation when intensity changes occur, such as in dynamic contrast‑enhanced (DCE) MRI.

\section{Acknowledgments}
This work was supported by the National Natural Science Foundation of China under Grant 62571328.

\bibliography{aaai2026}
\newpage

\section{Appendix}

\subsection{Details of Data Processing}
For both prospective and retrospective studies, coil sensitivity maps are estimated from the time‑averaged k‑space data using the ESPIRiT algorithm~\cite{uecker2014espirit}, implemented via the Python library \texttt{SigPy}~(\url{https://github.com/mikgroup/sigpy}).

\subsection{Additional Details of Baselines.}

\subsubsection{$\ell1$-Wavelet}
A compressed sensing method that exploits sparsity in the wavelet domain with an $\ell_1$ regularization.
In our experiments, we use the open‑source implementation from the Python library \texttt{SigPy}~(\url{https://github.com/mikgroup/sigpy}).

\subsubsection{GRASP}
A compressed sensing framework for golden‑angle radial sampling that jointly reconstructs images with temporal sparsity constraints.
We use the official MATLAB implementation provided by the authors
(\url{https://cai2r.net/resources/grasp-matlab-code/}.

\subsubsection{TDDIP}
A dynamic MRI reconstruction method based on the Deep Image Prior (DIP), which leverages an untrained convolutional neural network to exploit spatiotemporal redundancy. We use the official implementation provided by the authors (\url{https://github.com/jaejun-yoo/TDDIP/}.

\subsubsection{ST-INR (L\&S)}
A state‑of‑the‑art INR‑based method that models dynamic images as continuous spatiotemporal functions.
It uses hash‑grid encoding for fast convergence and incorporates low‑rank and sparsity constraints to maintain temporal consistency.
We use the official implementation released by the authors (\url{https://github.com/AMRI-Lab/INR_for_DynamicMRI}).

\subsection{Implementation Details of Our MoCo-INR}
\subsubsection{Network Architecture.}
The DVF network $\mathcal{F}_{\boldsymbol{\varPhi}}$ uses hash encoding with the parameters $N_\text{min}=2$, $T=2^{21}$, $L=10$, $F=4$ and $b=2$, while the canonical network $\mathcal{G}_{\boldsymbol{\varPsi}}$ is configured with $N_\text{min}=2$, $T=2^{21}$, $L=12$, $F=8$ and $b=2$. 
Both networks employ lightweight CNN decoders composed of three convolutional layers. The first two convolutional layers are followed by nonlinear activation functions, with $64$ filters of size of $3$, and the final convolutional layer outputs without activation.

\subsubsection{Training Details.}
We use Adam optimizer~\cite{kingma2017adam} with default hyperparameters to jointly optimize $\mathcal{F}_{\boldsymbol{\varPhi}}$ and $\mathcal{G}_{\boldsymbol{\varPsi}}$.
The training runs for 1200 iterations and adopts the proposed coarse‑to‑fine hash‑encoding learning strategy:
\begin{itemize}
    \item \textbf{Iterations 1-400}: train DVF network with levels 1-6 and canonical network with levels 1–8.
    \item \textbf{Iterations 400-800}: train DVF network with levels 4-8 and canonical network with levels 6-10.
    \item \textbf{Iterations 800-1200}: train DVF network with levels 6-10 and canonical network with levels 8-12.
    
\end{itemize}

\begin{figure}[ht]
\centering
\includegraphics[width=\linewidth]{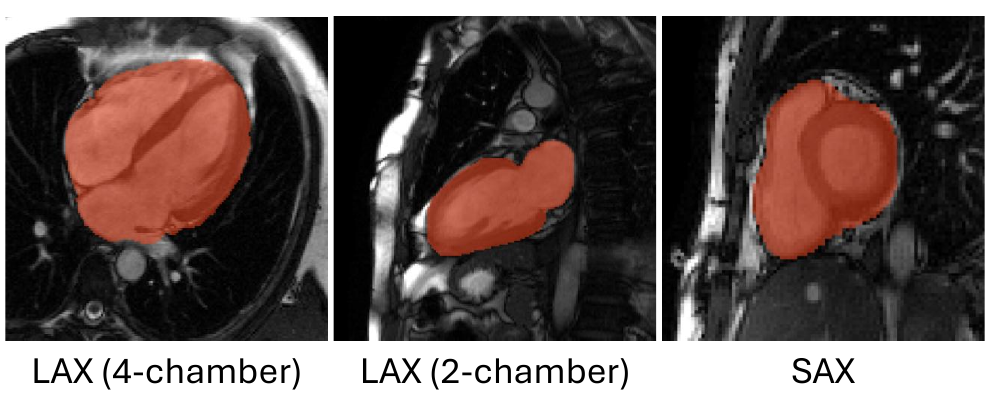} 
\caption{Illustration of the generated cardiac ROI masks for nRMSE (ROI) evaluation, where the orange region denotes the mask.}
\label{roi}
\end{figure}

\subsection{Cardiac ROI Mask Generation}
We use a two-step segmentation pipeline to generate our cardiac ROI region mask. First, the coarse segmentation labels are produced using MedSAM~\cite{ma2024segment}, guided by manual prompts. 
Second, an experienced radiologist manually refines the ROI masks to ensure anatomical accuracy.
Fig.~\ref{roi} illustrates the three cases of the segmented cardiac ROI region mask.

\subsection{Additional Visualization Results}
\begin{figure*}
    \centering
    \includegraphics[width=\textwidth]{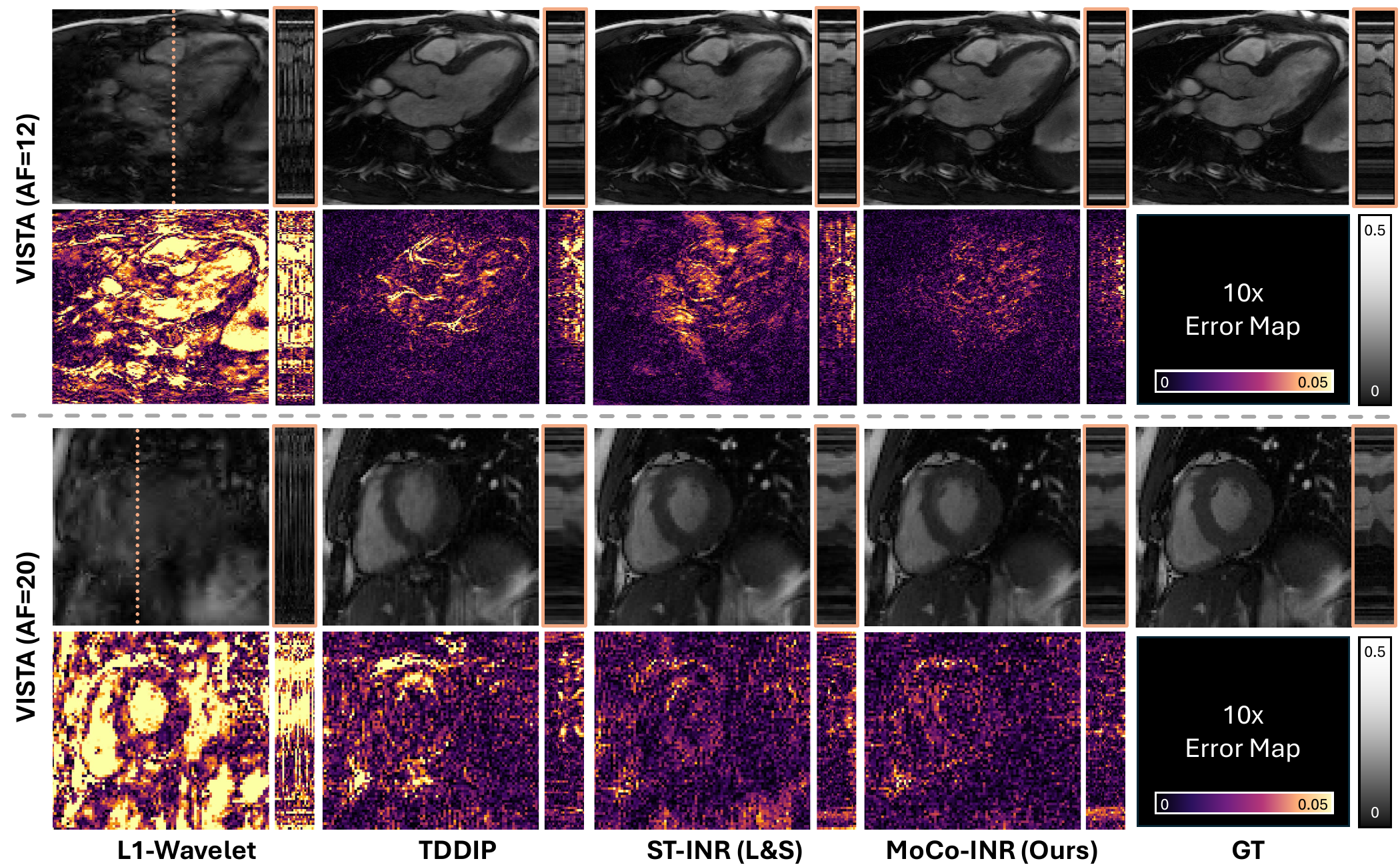}
    \caption{Qualitative comparisons of retrospective study on VISTA sampling for the baseline methods. }
    \label{fig:VISTA_recon}
\end{figure*}

\begin{figure*}
    \centering
    \includegraphics[width=\textwidth]{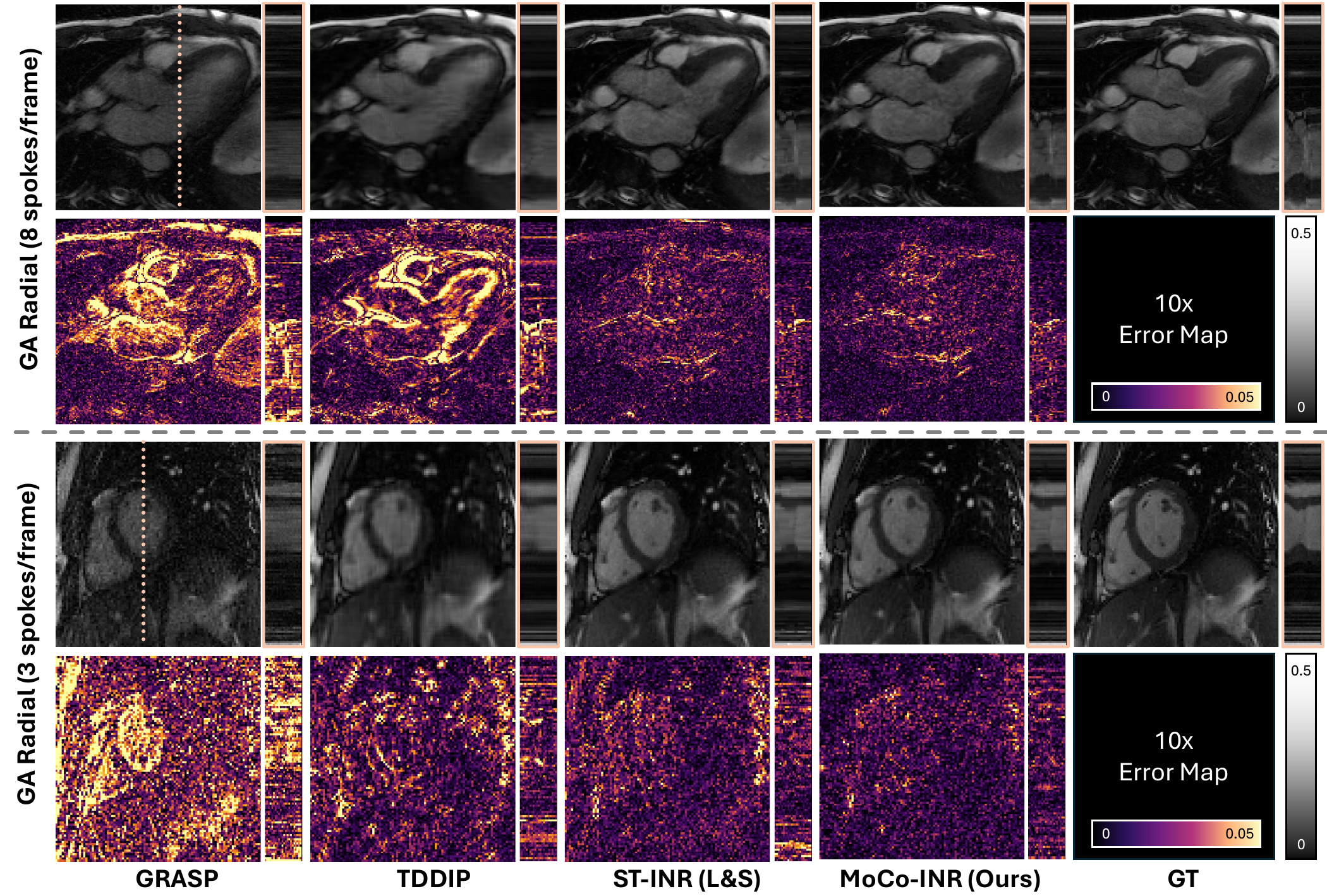}
    \caption{Qualitative comparisons of retrospective study on golden-angle (GA) radial sampling for the baseline methods.}
    \label{fig:GA_recon}
\end{figure*}

We provide a \textbf{reconstruction video}, including real free‑breathing acquisitions, to better illustrate the dynamic changes (please refer to the attached videos). 
Fig.~\ref{fig:VISTA_recon} and Fig.~\ref{fig:GA_recon} present additional retrospective reconstruction results on two sampling patterns, comparing MoCo‑INR with baseline methods.
MoCo‑INR achieves the best reconstruction quality, exhibiting the lowest reconstruction errors as shown in the corresponding error maps.

\end{document}